\documentclass[a4paper,11pt]{article}
\usepackage{jinstpub} 
\usepackage{lineno}
\usepackage{makecell}
\newcommand{\nnbar}{$n\rightarrow\bar{n}$ }

\title{\boldmath First application of a liquid argon time projection chamber for the search for intranuclear neutron-antineutron transitions  and annihilation in $^{40}$Ar using the MicroBooNE detector}

\collaboration{MicroBooNE Collaboration}

\author[jj]{P.~Abratenko}
\author[jj]{O.~Alterkait}
\author[o]{D.~Andrade~Aldana}
\author[t]{L.~Arellano}
\author[ii]{J.~Asaadi}
\author[gg]{A.~Ashkenazi}
\author[l]{S.~Balasubramanian}
\author[l]{B.~Baller}
\author[z]{G.~Barr}
\author[z]{D.~Barrow}
% ...

% ...
\affiliation[l]{Fermi National Accelerator Laboratory (FNAL), Batavia, IL 60510, USA}
% ...
\affiliation[o]{Illinois Institute of Technology (IIT), Chicago, IL 60616, USA}
% ...
\affiliation[t]{The University of Manchester, Manchester M13 9PL, United Kingdom}
% ...
\affiliation[z]{University of Oxford, Oxford OX1 3RH, United Kingdom}
% ...
\affiliation[gg]{Tel Aviv University, Tel Aviv, Israel, 69978}
% ...
\affiliation[ii]{University of Texas, Arlington, TX, 76019, USA}
\affiliation[jj]{Tufts University, Medford, MA, 02155, USA}
% ...

  \emailAdd{microboone\_info@fnal.gov}
\date{}

\abstract{We present a novel methodology to search for intranuclear neutron-antineutron transition ($n\rightarrow\bar{n}$) followed by 
$\bar{n}$-nucleon annihilation within an $^{40}$Ar nucleus, using the MicroBooNE liquid argon time projection chamber (LArTPC) detector. A discovery of $n\rightarrow\bar{n}$ transition or a new best limit on the lifetime of this process would either constitute physics beyond the Standard Model or greatly constrain theories of baryogenesis, respectively. The approach presented in this paper makes use of deep learning methods to select $n\rightarrow\bar{n}$ events based on their unique features and differentiate them from cosmogenic backgrounds. The achieved signal and background efficiencies are (70.22$\pm$6.04)\% and (0.0020$\pm$0.0003)\%, respectively. A demonstration of a search is performed with a data set corresponding to an exposure of $3.32 \times10^{26}\,$neutron-years, and where the background rate is constrained through direct measurement, assuming the presence of a negligible signal. With this approach, no excess of events over the background prediction is observed, setting a demonstrative lower bound on the $n\rightarrow\bar{n}$ lifetime in $^{40}$Ar of $\tau_{\textrm{m}} \gtrsim 1.1\times10^{26}\,$years, and on the free $n\rightarrow\bar{n}$ transition time of $\tau_{\textrm{\nnbar}} \gtrsim 2.6\times10^{5}\,$s, each at the $90\%$ confidence level. This analysis represents a first-ever proof-of-principle demonstration of the ability to search for this rare process in LArTPCs with high efficiency and low background.}


\arxivnumber{2308.03924} 

\begin{document}{
\maketitle
\flushbottom

\section{Introduction}
\label{intro}
Processes such as neutron-antineutron transition~\cite{nnbar:2013} can provide a unique test of theoretical \hbox{extensions} to the Standard Model of particle physics that allow for the violation of baryon number conservation~\cite{SM}. The transition of a neutron to antineutron ($n\rightarrow\bar{n}$) is a theoretically motivated beyond-Standard Model process which violates baryon number by two units~\cite{nnbar:2012, intro1Nussinov:2001rb, intro3Mohapatra:1980qe, intro4Addazi:2020nlz, intro5Phillips:2014fgb}. The process of intranuclear $n\rightarrow\bar{n}$ involves the transformation of a bound neutron into an antineutron. This antineutron then annihilates with a nearby nucleon (neutron or proton) and produces, on average, $3$--$4$ final state pions~\cite{PhysRevD.103.012008,Golubeva:2018mrz}. The branching ratios of $\bar{n}p$ and $\bar{n}n$ annihilation products are based on past measurements of $\bar{p}n$ and $\bar{p}p$ interactions, respectively~\cite{PhysRevD.103.012008,Golubeva:2018mrz,Klempt_2005,AMSLER2003357}. In a vacuum, the final state pions produced by a motionless and unbound annihilating pair are expected to have zero total momentum and a total invariant mass corresponding to the sum of the masses of the two (anti)nucleons. Deviations from this expectation are due to nuclear effects---specifically, intranuclear Fermi motion of the annihilating (anti)nucleons, their nuclear binding energy, and final state interactions as the initial state mesons traverse the nuclear medium---leading to smearing effects of the observed final state kinematics. The annihilation has a star-like, spherical topological signature, which can be used to differentiate it from background interactions.

An experimental discovery or stringent lower bound, surpassing the current best limits~\cite{PhysRevD.103.012008,Baldo-Ceolin:1994hzw}, on the rate of intranuclear $n\rightarrow\bar{n}$ would make an important contribution to our understanding of the baryon asymmetry of the Universe. To date, limits have been placed on the mean lifetime of this process by various experiments using either free neutrons or neutrons bound in nuclei~\cite{1983_homestake, 1984_IMB, 1985_ILL, kamiokande, Triga, Frejus, 2015_SK, ILL, 2002_soudan2, 2017_SNO}. The free-neutron $n\rightarrow\bar{n}$ lifetime ($\tau_{\textrm{\nnbar}}$) and bound-neutron $n\rightarrow\bar{n}$ lifetime ($\tau_{\textrm{m}}$) are related through a factor ($R$)~\cite{Barrow:2019viz,PhysRevD.78.016002,Barrow:2021svu} as shown in eq. (\ref{eq:lifecor}), which accounts for the high suppression of the transition due to differences in the nuclear potentials of neutrons and antineutrons within the nucleus where this process could take place,
\begin{equation}
\tau_{\textrm{m}}=R\tau_{\textrm{\nnbar}}^{2}.
\label{eq:lifecor}
\end{equation}
For $^{40}$Ar nuclei, $R$ is expected to take on a value of $5.6\times 10^{22}\,$s$^{-1}$ with an uncertainty of $20\%$~\cite{Barrow:2019viz}. The most stringent limit on the free neutron transition time is provided by ILL in Grenoble~\cite{Baldo-Ceolin:1994hzw} at $0.86\times 10^8\,$s at the $90\%$ confidence level (CL), while the Super-Kamiokande experiment~\cite{2015_SK}, using oxygen-bound neutrons and an associated suppression factor of $5.17\times 10^{22}\,$s$^{-1}$~\cite{PhysRevD.78.016002,Barrow:2021svu}, corresponds to $\tau_{\textrm \nnbar}>4.7\times10^8\,$s at the $90\%$ CL~\cite{PhysRevD.103.012008}.

The future Deep Underground Neutrino Experiment (DUNE) will be able to provide competitive limits on the lifetime of this process because of its much larger detector mass~\cite{DUNE1:2016oaz,DUNE2:2016oaz,DUNE3:2016oaz}. The preliminary simulation studies from DUNE~\cite{Hewes:2017xtr} show the ability of deep-learning (DL) networks when combined with high-resolution images from liquid argon time projection chamber (LArTPC) detectors to extend the sensitivity reach for \nnbar process. However, the DUNE simulation studies do not account for detector mismodeling, a simplified assumption that is not the reality. Therefore, it is crucial to develop and validate DL-based techniques to select $n\rightarrow\bar{n}$ like signals using real LArTPC data. 

This work presents a DL-based analysis of MicroBooNE data, making use of a sparse convolutional neural network (CNN)~\cite{sCNN:2017,sCNN2:2017}, to search for $n\rightarrow\bar{n}$ like signals using primarily their topological signature. The results reported in this paper use the MicroBooNE off-beam data (data collected when the neutrino beam was not running) with a total exposure of $372\,$s corresponding to $3.32 \times10^{26}\,$neutron-years. The limited exposure is attributed to the design of the MicroBooNE detector, whose primary requirement was the study of neutrinos from a pulsed accelerator beam, thus restricting the data collection to short periods of time associated with beam and other external triggers.

\section{Experimental setup}\label{exp}
The MicroBooNE LArTPC detector~\cite{Acciarri:2016smi} employs an active volume of 85 metric tonnes of liquid argon (LAr). The detector is a $10.4\,$m long, $2.6\,$m wide, and $2.3\,$m high LArTPC and is located on-surface and on-axis to the Booster Neutrino Beamline~\cite{BNB} at Fermilab. Due to its on-surface location, the MicroBooNE detector is exposed to a large flux of cosmic rays, leading to a variety of cosmogenic activity in the detector. Charged particles produced from interactions within the LAr leave a trail of ionization electrons which drift, under the effect of a uniform electric field, with a maximum electron drift time (time taken by ionization electrons to drift to the anode wires) of $2.3\,$ms towards anode wire planes. Three anode wire planes named $U$, $V$, and $Y$, with $U$ and $V$ plane wires oriented at $\pm60^{\circ}$ relative to vertical, and $Y$ plane wires oriented vertically, sense and collect the ionization charge. A light detection system composed of photomultiplier tubes (PMT) detects scintillation light produced in the interaction which in turn helps to determine the drift time, achieving 3D particle reconstruction. Data was collected from 2015--2021 and includes off-beam data during periods when there was no neutrino beam.
\section{Analysis overview}
\label{sec:anaov}
The methodology used to search for a \nnbar intranuclear transition in MicroBooNE was developed using off-beam data that were recorded using an external, random trigger. Each trigger corresponds to an exposure of $2.3\,$ms (an ``event''), the standard readout length of MicroBooNE. The readout window (or exposure interval) ensures that all ionization information associated with a given interaction at trigger time occurring anywhere in the active volume is collected by the readout. During this period, light and unbiased (raw) ionization charge data were collected and analyzed, searching for interaction ``clusters'' with a characteristic star-like topology. The dominant source of interactions during these short beam-off exposures comes from cosmic ray muons (straight track-like features) and other cosmogenic activity, and/or products of their electromagnetic and hadronic showers, which are expected to contribute as the dominant background to this $n\rightarrow\bar{n}$ search. This source of background is a unique issue to a search using the MicroBooNE detector, due to its on-surface location, whereas searches with detectors located deep underground, such as DUNE, are expected to be limited by atmospheric neutrino backgrounds~\cite{DUNE2:2016oaz}.
\subsection{Data-driven background}
\label{sec:bkgsim}
MicroBooNE does not use a dedicated Monte Carlo simulation for cosmic backgrounds (which includes any activity produced by primary cosmic muons, cosmic neutrons, cosmic antineutrons, and cosmic antiprotons) but instead relies on in situ measurements to directly measure and thus constrain the rate of these interactions as backgrounds to beam-related analyses. As such, a data-driven approach was followed to search for $n\rightarrow\bar{n}$ under the assumption of negligible signal being present in the data. In this approach, the off-beam data sample was divided into four statistically independent sub-samples, where $40\%$ was reserved for analysis development and, in particular, to train machine learning algorithms, $50\%$ was reserved as the test sample to determine signal selection efficiency and predict background rates, $5\%$ was set aside for the development validation of a blinded analysis using ``fake data'', and the remaining $5\%$ corresponding to $372\,$s of exposure was reserved as the ``data'' sample for the final measurement and reported results. This analysis was performed blind, with final data distributions and extracted $n\rightarrow\bar{n}$ limits obtained only after the review of the analysis and its validation using fake data as will be described in section~\ref{sec:fakedata}. The data-driven approach used to generate the signal and background samples automatically enables accurate ``modeling'' of cosmogenic activity and noise sources, including any time dependence in the detector response. However, this approach assumes that there are $<10^{-6}$ \nnbar interaction events in the 
 off-beam data corresponding to $372\,$s of exposure. This is a safe assumption given the current best limits on $n\rightarrow\bar{n}$ from the Super-Kamiokande experiment~\cite{PhysRevD.103.012008} and MicroBooNE's low exposure as explained in section~\ref{intro}. 
\subsection{Signal simulation}
\label{sec:sigsim}
 Signal $n\rightarrow\bar{n}$ interactions are simulated across the detector's active volume using the GENIE neutrino event generator (GENIE v.3.00.04)~\cite{Andreopoulos:2015wxa,Hewes:2017xtr}. These interactions are simulated with annihilation vertices uniformly distributed across the active volume of the TPC and as a consequence, a significant fraction of the signal interactions are only partially visible in the TPC. This leads to inefficiencies which are accounted for in the reported signal efficiencies which are shown in table~\ref{tab:test_combined_efficiency}. In the model, the (anti)nucleon's Fermi motion and binding energy are modeled using a local Fermi gas model, and the empirical, data-driven hA Intranuke algorithm is used to simulate final state interactions (FSI). The $^{40}$Ar nucleus is assumed to be at rest during the \nnbar process. The position of a neutron (to be oscillated into an antineutron) within the nucleus is simulated using GENIE's density profile of nucleons (Woods-Saxon distribution~\cite{WoodsSaxon}),
\begin{equation}
    \rho(r) = \frac{\rho_{0}}{1+e^{\frac{r-R_{0}}{a}}}~,
\end{equation}
where $r$ is the radial position inside the nucleus, $R_0 = r_{0}A^{\frac{1}{3}}$ is the nuclear radius, with $r_{0}$ defined as $1.4\,$fm in GENIE. $\rho_{0}$ is normalized in order to express nuclear density as a probability distribution, and $a$ is a parameter describing the surface thickness of the nucleus, set to $a=0.54$~fm.
 
This analysis considers the annihilation of an antineutron with either a neutron or a proton and simulates the resulting products of annihilation ($3$--$4$ pions on an average) using the branching ratios informed by previous measurements~\cite{PhysRevD.103.012008,Klempt_2005,AMSLER2003357}, reproduced in table~\ref{tab:branching_ratio}, accounting for the available kinematic phase-space on an event-by-event basis~\cite{Hewes:2017xtr}.  
 The final state particles are subsequently propagated through the detector with Geant4~\cite{Geant4}. This is followed by the custom detector simulation for the MicroBooNE detector~\cite{Snider:2017wjd, Pordes:2016ycs, g4_hA} to take account of the detector response. 
 
\begin{table}[htbp]
\caption{\label{tab:branching_ratio}Effective branching ratios for antineutron annihilation in $^{40}$Ar, as implemented in GENIE. The branching ratios are adapted from analysis by the Super-Kamiokande collaboration~\cite{PhysRevD.103.012008} and are derived from past antiproton annihilation measurements on hydrogen and deuterium, with a phase-space approximation ~\cite{Hewes:2017xtr}.}
    \centering
    \begin{tabular}{cc|cc}
    \hline\hline
    \multicolumn{2}{c}{$\bar{n}+p$} & \multicolumn{2}{c}{$\bar{n}+n$}\\
 \hline
         Channel & Branching ratio & Channel & Branching ratio \\
         \hline\hline
         $\pi^{+}\pi^{0}$ & 1.2\% & $\pi^{+}\pi^{-}$ & 2.0\% \\
         $\pi^{+}2\pi^{0}$ & 9.5\% & $2\pi^{0}$ & 1.5\% \\
         $\pi^{+}3\pi^{0}$ & 11.9\% & $\pi^{+}\pi^{-}\pi^{0}$ & 6.5\% \\
         $2\pi^{+}\pi^{-}\pi^{0}$ & 26.2\% & $\pi^{+}\pi^{-}2\pi^{0}$ & 11.0\% \\
         $2\pi^{+}\pi^{-}2\pi^{0}$ & 42.8\% & $\pi^{+}\pi^{-}3\pi^{0}$ & 28.0\% \\
         $2\pi^{+}\pi^{-}2\omega$ & 0.003\% & $2\pi^{+}2\pi^{-}$ & 7.1\% \\
         $3\pi^{+}2\pi^{-}\pi^{0}$ & 8.4\% & $2\pi^{+}2\pi^{-}\pi^{0}$ & 24.0\% \\
          &  & $\pi^{+}\pi^{-}\omega$ & 10.0\% \\
          &  & $2\pi^{+}2\pi^{-}2\pi^{0}$ & 10.0\% \\
\hline\hline
    \end{tabular}
   
\end{table}

\subsection{Overlay generation}
\label{sec:ovrlysim}
Neutron-antineutron transition signal interactions are simulated within GENIE as described in section~\ref{sec:sigsim} and are then overlaid onto the background (real cosmic data) at waveform level to emulate the events used to estimate the effective signal efficiency. An example of overlay generation is shown in figure~\ref{fig:overlay} where the top panel shows the background event and the bottom panel shows the overlay scenario where the GENIE simulated signal interaction, circled in red, is overlaid on the same background event shown in the top panel.

Because of abundant cosmogenic activity, each $2.3\,$ms event includes multiple reconstructed cosmic candidate interactions in the LAr volume, referred to as ``clusters''. Three-dimensional clusters are reconstructed using the WireCell reconstruction package~\cite{WC:JINST} as collections of $3$D spacepoints, where each spacepoint carries information about its corresponding wire position, time-tick, and charge deposition. The true $n\rightarrow\bar{n}$ interaction clusters are identified through the comparison of two events (one with and one without a signal interaction) with the same background source, as depicted in figure~\ref{fig:overlay}. The topological features of the signal clusters (``star-like'') and the background clusters (``straight track-like'') are then used to develop the selection as described in the next section.  


\begin{figure*}[htbp]
\centering 
\includegraphics[width=1.0\textwidth]{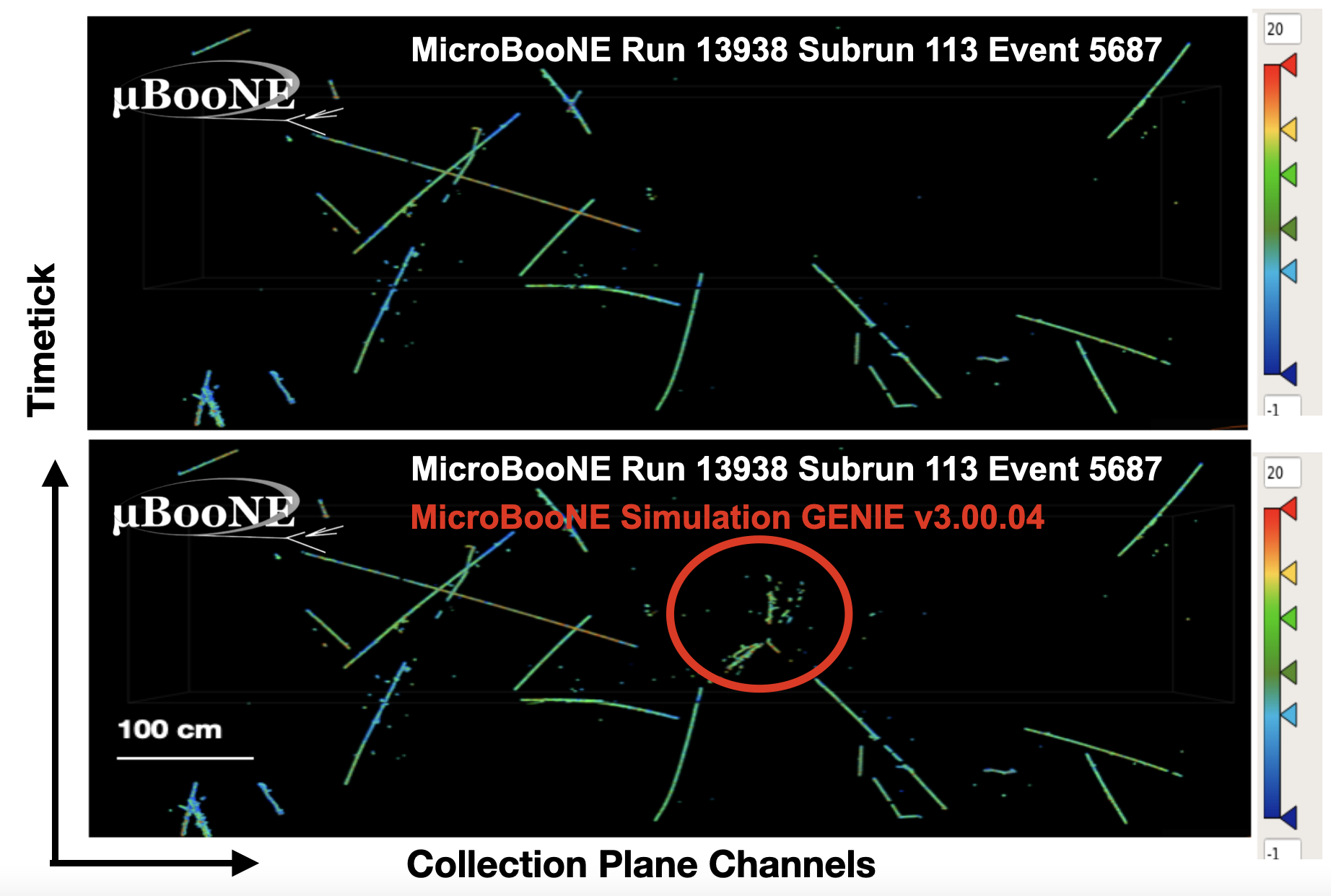}
\caption{
(top) Event display showing an event collected in MicroBooNE off-beam data. (bottom) Event display showing the same off-beam data event with a GENIE-simulated \nnbar signal cluster (highlighted in the red circle). The vertical and horizontal scales are the same. The $x$-axis represents the collection plane channels and the $y$-axis represents the time-tick space. Color represents the amount of deposited ionization charge in units of ADC values.
}
\label{fig:overlay}
\end{figure*}

\section{Analysis techniques and selection criteria}\label{sel}
The cluster reconstruction is followed by a series of selection criteria which are applied in three stages as discussed in the following subsections.
\subsection{BDT-based preselection}
\label{presel}
The first, or preselection, stage makes use of a machine learning boosted decision tree (BDT) algorithm using xgboost~\cite{xgbdt} to significantly reduce the number of background clusters while maintaining high signal efficiency. The BDT is trained using variables that contain information about the number of spacepoints in each cluster, with their wire positions and times. The distributions of these input variables are shown in figure~\ref{fig:BDT_input}, corresponding to the ``No selection'' stage of table~\ref{tab:test_combined_efficiency}. We define the ``extent'' of a cluster as the number of wires or time-ticks over which the cluster is contained in the $U$, $V$, or $Y$ wire-plane or time-tick dimension (one time-tick corresponds to $0.5$ $\mu$s), respectively. These variables enable us to distinguish between signal and background clusters based on their topological features, such as the more localized, spherical topology for the signal \nnbar clusters and the straight, track-like topology for the background clusters; effective in significantly suppressing the cosmic muon backgrounds

\begin{figure*}[htbp]
\centering
\includegraphics[width=0.4\linewidth]{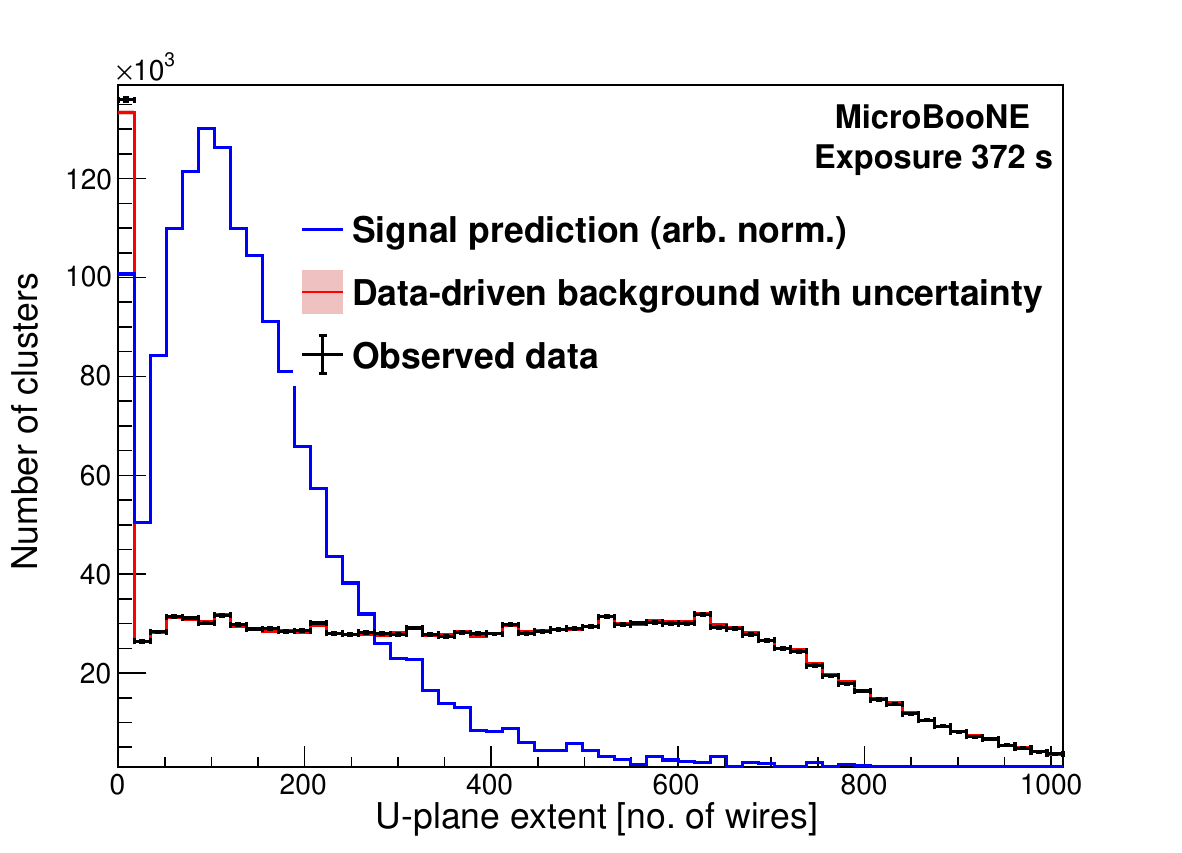}
\includegraphics[width=0.4\linewidth]{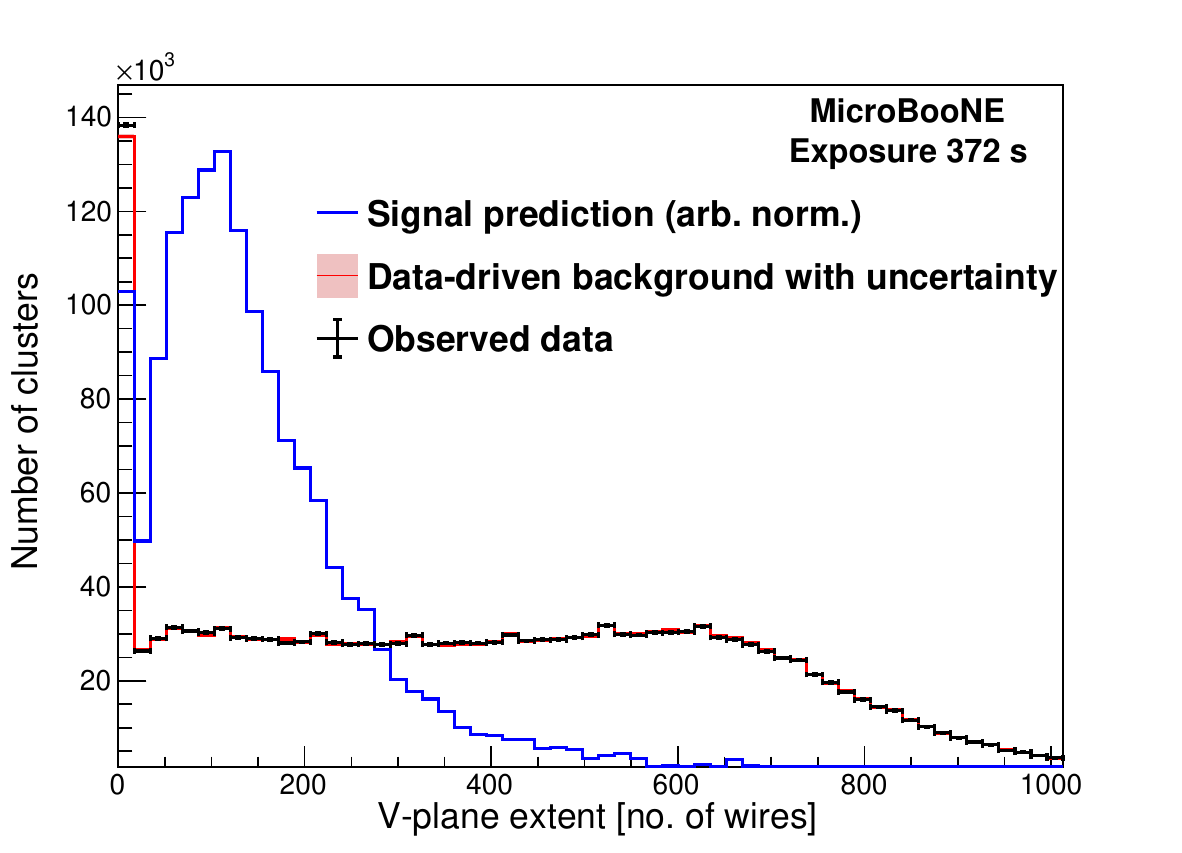}
\newline
\includegraphics[width=0.4\linewidth]{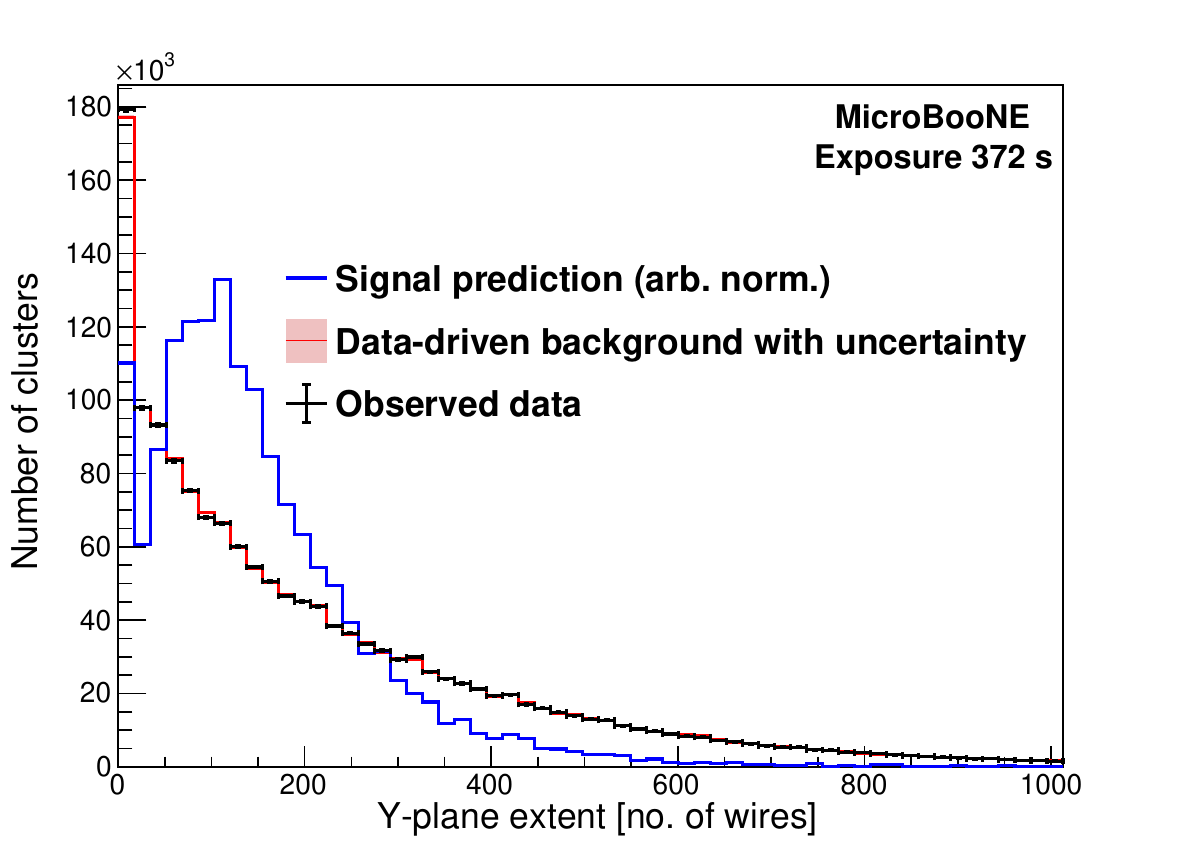}
\includegraphics[width=0.4\linewidth]{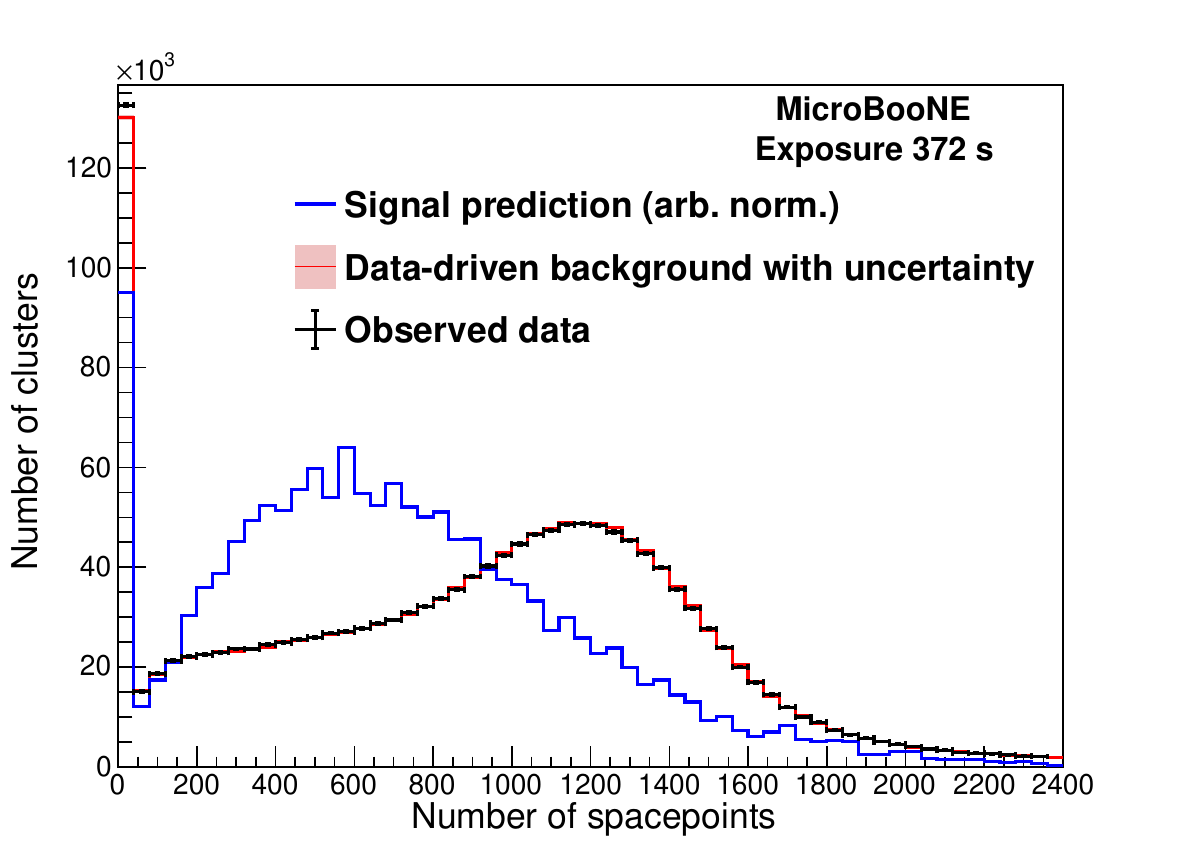}
\newline
\includegraphics[width=0.4\linewidth]{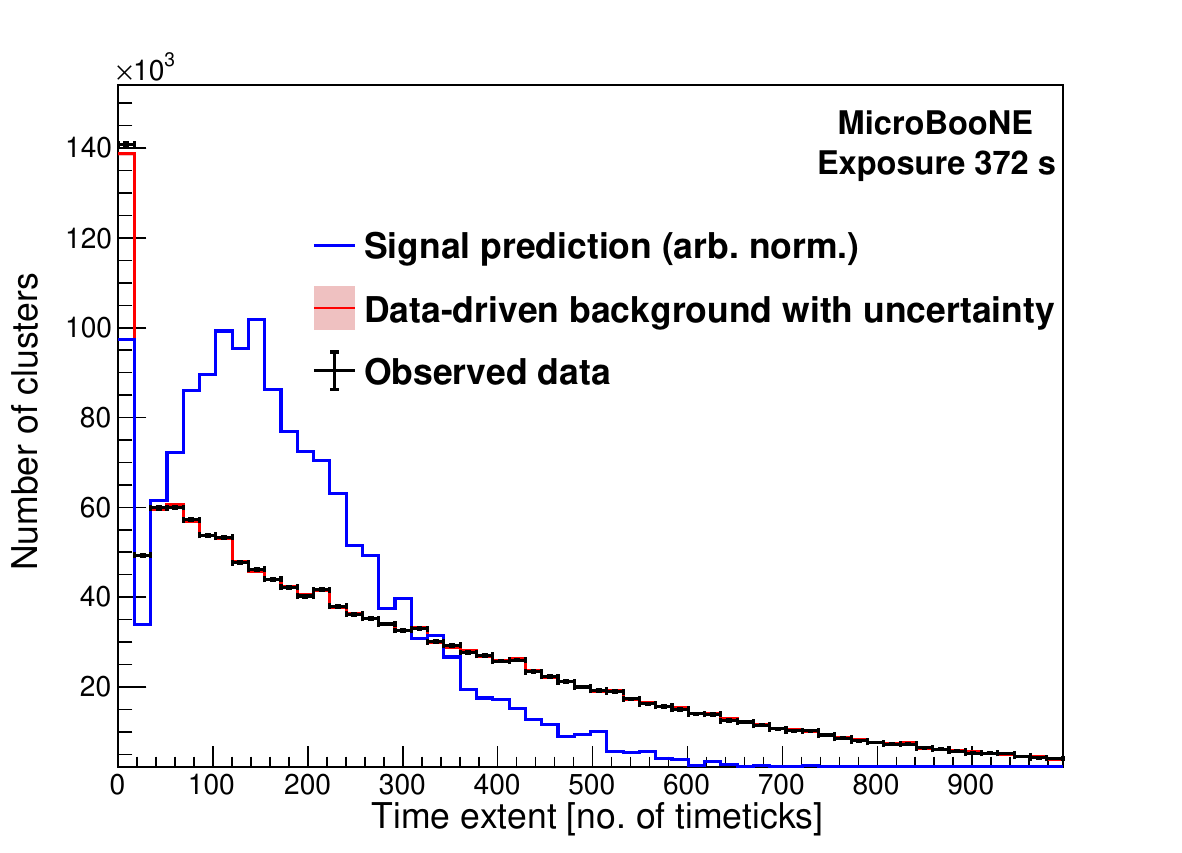}
\caption{Distributions of topological variables from 2D cluster projections for the signal (blue) and background (red) clusters. The background is shown along with its systematic uncertainty band (see section~\ref{sec:uncs} for details on assessing systematic uncertainties). The systematic uncertainty is small and of the order of a few percent. The data points corresponding to $372\,$s of exposure are shown (after unblinding) in black along with statistical uncertainty. The background clusters, generated with a test sample, are normalized exactly to match the data exposure of $372\,$s, whereas the signal clusters, which were simulated and overlaid onto the background clusters, are arbitrarily normalized as they cannot be precisely scaled to match the data exposure. The samples used to obtain background prediction and data are assumed to have a negligible signal.}
\label{fig:BDT_input}
\end{figure*}

The BDT training outcome exhibits a clear separation between the signal ($n\rightarrow\bar{n}$) and background (cosmic) processes, as shown in figure~\ref{fig:bdttrainingResults} (the BDT score distribution corresponds to the ``No selection'' stage of table~\ref{tab:test_combined_efficiency}.). Therefore, a ``loose'' cut on the BDT score is chosen by visual inspection to meet our primary goal of rejecting the majority of background events at this stage. Selecting clusters with BDT score $>0.1$ (see figure~\ref{fig:bdttrainingResults}) rejects $91\%$ of the background clusters and maintains a high signal selection efficiency of $86\%$.

\begin{figure*}[htbp]
    
\centering 
\includegraphics[width=.49\textwidth]{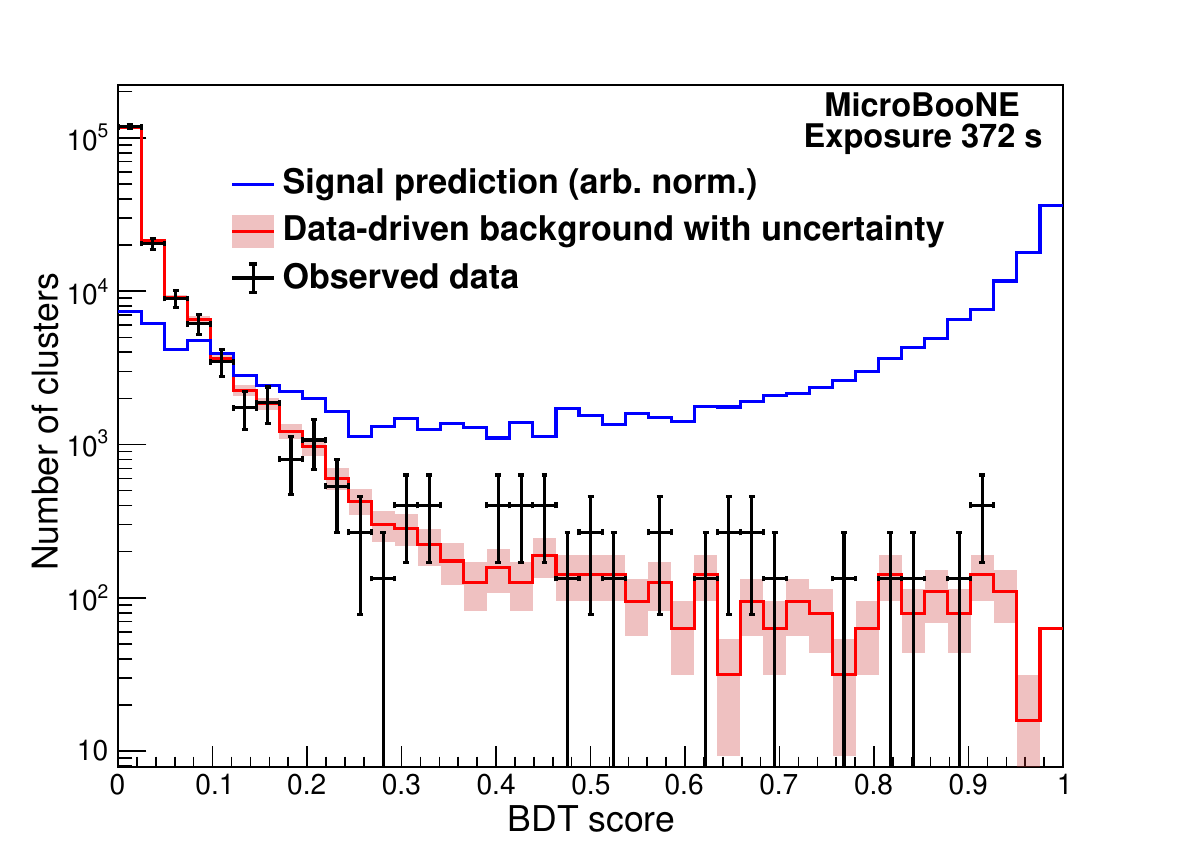}
\includegraphics[width=.49\textwidth]{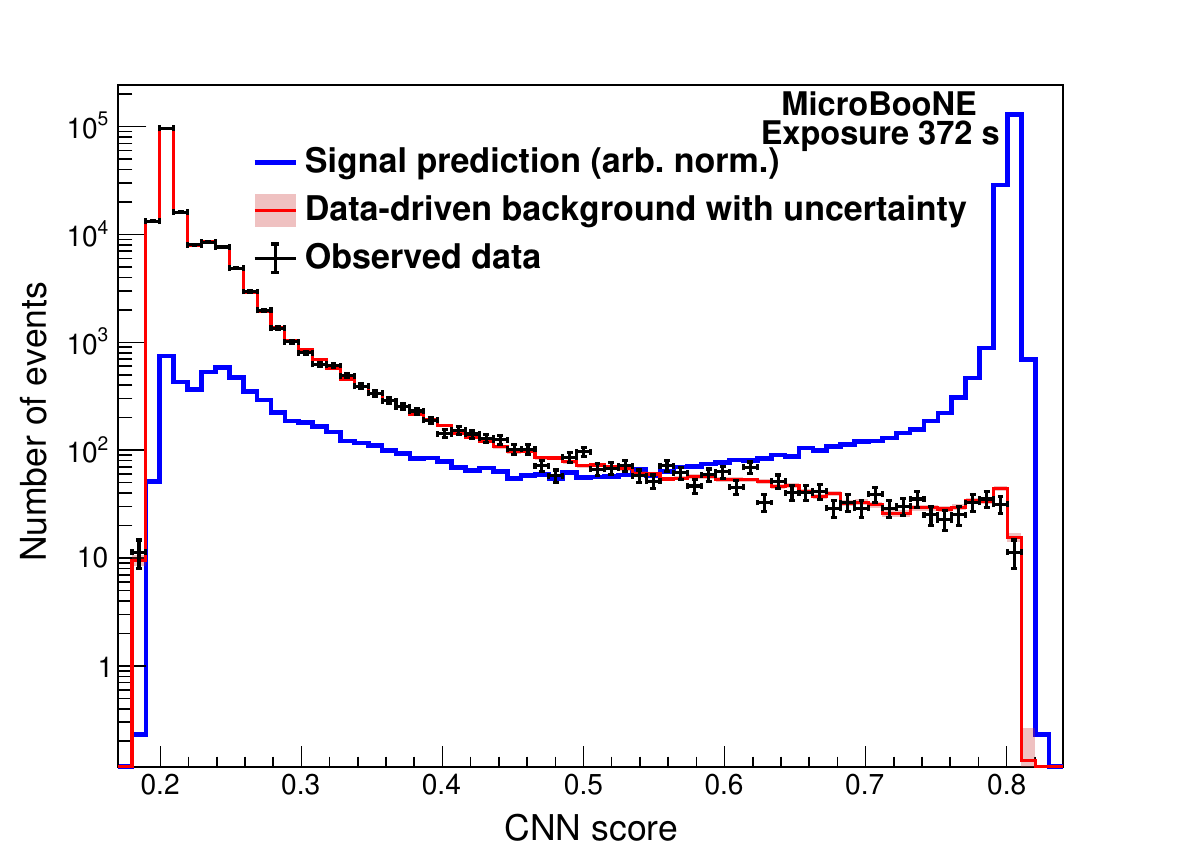}
\caption{
Classification performance of the BDT (left) and CNN (right) for the signal $n\rightarrow\bar{n}$ (blue) and background (red) clusters. The background is shown along with the systematic uncertainty band (see section ~\ref{sec:uncs} for details on assessing systematic uncertainties). The data points corresponding to $372\,$s of exposure are shown (after unblinding) in black along with statistical uncertainty. The background clusters, generated with a test sample, are normalized exactly to match the data exposure of $372\,$s, whereas the signal clusters, which were simulated and overlaid onto the background clusters, are arbitrarily normalized as they cannot be precisely scaled to match the data exposure. The samples used to obtain background prediction and data are assumed to have a negligible signal.}
\label{fig:bdttrainingResults}
\end{figure*}
\subsection{CNN-based selection}
\label{cnnsel}
The second stage of selection applies an image-based selection criterion, using a sparse CNN with the VGG16 network architecture~\cite{sCNN:2017,sCNN2:2017,sCNN:ub2021,sCNN:uB2022}. Convolutional neural networks perform successive layers of convolutions on full images to identify features and associate these features with labels~\cite{MicroBooNE:2020hho}. A sparse CNN makes use of localized inputs within an image (star-like topology for the signal clusters and straight track-like topology for the background clusters) that highlight features on which the network trains rather than the full image. This selection stage makes use of 2D projections of the preselected clusters onto three sense wire planes of the MicroBooNE detector. These projections contain information about the wire position, time-tick, and charge deposition associated with each cluster, and are formatted in such a way as to retain only the pixels associated with the signal or background clusters, thus making it highly memory efficient. 

The network is trained using nearly a million events. One challenge in training arises from having a finite dataset, leading to the possible risk of overtraining (or overfitting). In overtraining, the network learns a combination of features associated with a specific type of event and may occasionally incorporate features that are not representative of the broader event type. Therefore, it is crucial to validate the effectiveness of the network, which in this case is achieved by monitoring training loss and validation accuracy. The training loss represents the classification error in the training set and a decreasing training loss typically indicates that the model is learning to better fit the training data. The validation accuracy measures the performance of a model on a separate dataset that the model has not been trained on. An increasing validation accuracy indicates that the model is generalizing well to new data, whereas a decreasing validation accuracy may suggest overtraining. Figure~\ref{fig:cnntraining} shows the training loss and validation accuracy versus number of iterations which refer to the number of updates made to the model's parameters during the training. For CNN-based classification, 20,000 iterations were chosen by monitoring training loss and validation accuracy. The performance of the trained CNN on the test sample is shown in figure~\ref{fig:bdttrainingResults} (the right CNN score distribution corresponds to the ``Stage 1'' of table~\ref{tab:test_combined_efficiency}). 
\begin{figure*}[htbp]
    
\centering 
\includegraphics[width=1.\textwidth]{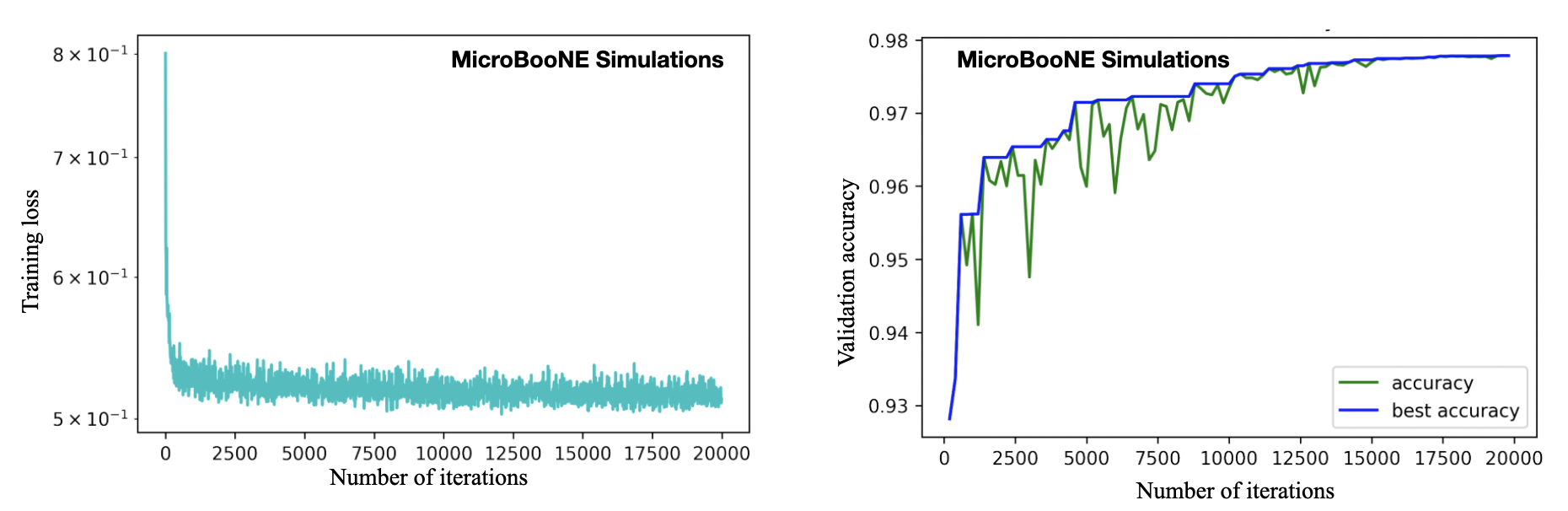}

\caption{Distributions of training loss (left) and validation accuracy (right) versus number of iterations. The current accuracy is shown as green solid line, whereas the best accuracy is shown as blue solid line. The best accuracy represents the highest validation accuracy observed throughout the training process.}
\label{fig:cnntraining}
\end{figure*}

The CNN score criterion is optimized with respect to the projected sensitivity at 90\% CL. As a prerequisite for the sensitivity calculation, efficiencies for the signal and background events are calculated for various CNN score criteria and are shown in table~\ref{tab:CNNopt}. For these particular CNN score criteria (where the background rejection is $\sim99\%$), individual preliminary sensitivity values were calculated, using the TRolke statistical method in ROOT~\cite{Rolke} following a frequentist approach and accounting for both statistical and systematic uncertainties on the background and signal efficiency, based on the following assumptions:
\begin{itemize}{
    \item The assumed search region statistics correspond to $372\,$s of exposure, and are evaluated by scaling the test sample (containing $\times10\,$ higher statistics) by a factor of 0.1, making it equivalent in size to the MicroBooNE ``data'' statistics for the analysis.
    \item The statistical uncertainty on the background is considered within the TRolke method, assuming Gaussian fluctuations on the data-sized test. The sensitivity calculation within TRolke assumes zero signal and hence no statistical uncertainty is assumed on the signal.
    \item For the CNN score criterion optimization study, the systematic uncertainty on the signal selection efficiency was assumed to be $15\%$. The systematic uncertainty on the background is evaluated as the statistical uncertainty on the background obtained using the test sample, as the background is measured in situ}.

\end{itemize}
Considering sensitivity (see table~\ref{tab:CNNopt}) as a figure of merit, the optimal CNN criterion is found to be $0.80$.

\begin{table}[htbp]
\centering
\caption{\label{tab:CNNopt}Preliminary sensitivity for various CNN score criteria around the optimized score of $0.80$. The signal and background efficiencies are calculated using the test sample. The background is also estimated using the test sample and then scaled by a factor of $0.1$ to make it equivalent in size to the MicroBooNE data sample which corresponds to $372\,$s of exposure. Uncertainties in the table account for finite MC statistics only.
}
\smallskip
\begin{tabular}{l|c|c|c|c}
\hline
 \thead{CNN criterion\\} & \thead{Signal Efficiency} & \thead{Background Efficiency\\($10^{-4}$)} & \thead{Normalized Background\\ Estimate} & \thead{Sensitivity\\($10^{25}$ yrs)} \\

    \hline
    0.797 & 0.8274 $\pm$ 0.0003 & 1.53 $\pm$ 0.10  & 24.8 $\pm$ 1.6 & 2.62\\
    0.798 & 0.8222 $\pm$ 0.0003 & 1.27 $\pm$ 0.09  & 20.5 $\pm$ 1.4 & 2.83\\
    0.799 & 0.8012 $\pm$ 0.0003 & 1.08 $\pm$ 0.08  & 17.5 $\pm$ 1.3 & 2.98\\
    0.800 & 0.7360 $\pm$ 0.0003 & 0.88 $\pm$ 0.07  & 14.2 $\pm$ 1.2 & 2.99\\
    0.801 & 0.6392 $\pm$ 0.0004 & 0.66 $\pm$ 0.06 & 10.7 $\pm$ 1.0 & 2.95\\
    0.802 & 0.5081 $\pm$ 0.0004 & 0.50 $\pm$ 0.06  & 8.1 $\pm$ 0.9 & 2.65\\
    0.803 & 0.3490 $\pm$ 0.0004 & 0.43 $\pm$ 0.05 & 6.9 $\pm$ 0.8 & 1.95\\
\hline
    \end{tabular}
\end{table}
\subsection{Topological-based final selection}
After CNN selection, approximately $2\%$ of the remaining clusters have zero extent in time or one of the wire dimensions, as a consequence of reconstruction inefficiencies~\cite{PhysRevLett.128.151801}. Those clusters looked similar before the preselection stage (as shown in figure~\ref{fig:BDT_input}), causing the BDT to be unable to distinguish between them. Therefore, a third and final selection stage, based on topological information, is applied to reject zero- and low-extent clusters, which cannot represent the signal topology. The distributions of extent variables after CNN selection (``Stage 2'' of table~\ref{tab:test_combined_efficiency}) are shown in figure~\ref{fig:1d_uextent} and the final selection criteria are chosen by visual inspection of these variables. The final selection requires the extent of a cluster in at least one of the three wire dimensions to be \hbox{$>70$} wires, and in the time dimension to be \hbox{$>70$} time-ticks. The final selection criteria were chosen to effectively reject the majority of background events, particularly those peaking in the range between 0 and 70 in extent as shown in figure~\ref{fig:1d_uextent}.\\
The number of signal and background events in the test sample before and after each of the three selection stages is shown in table ~\ref{tab:test_combined_efficiency}. The analysis yields an overall signal selection efficiency of $70.22\%$, corresponding to the ratio of events at Stage $3$ to events before any selection. At the same time, it rejects $99.99\%$ of the total background resulting in background efficiency of $0.0020\%$

\begin{figure*}[htbp]
\centering 
\includegraphics[width=.49\textwidth]{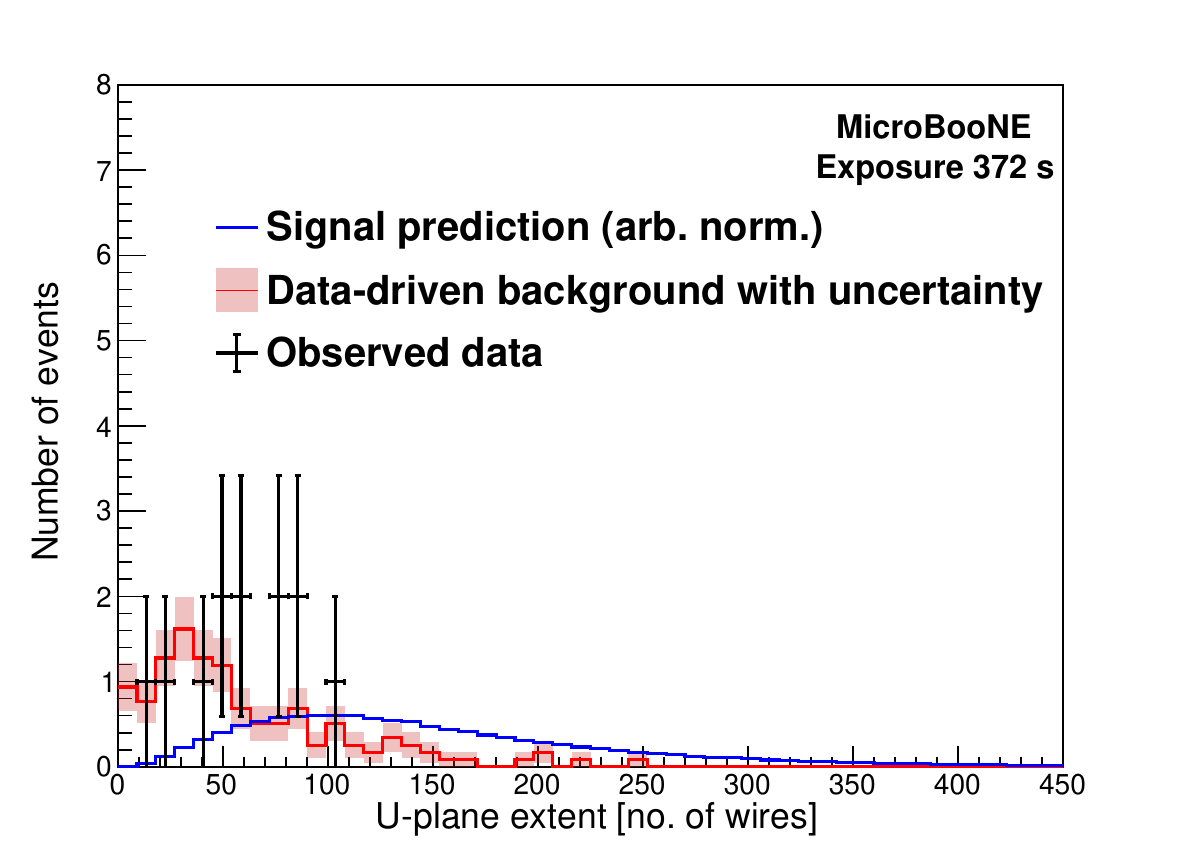}
\includegraphics[width=.49\textwidth]{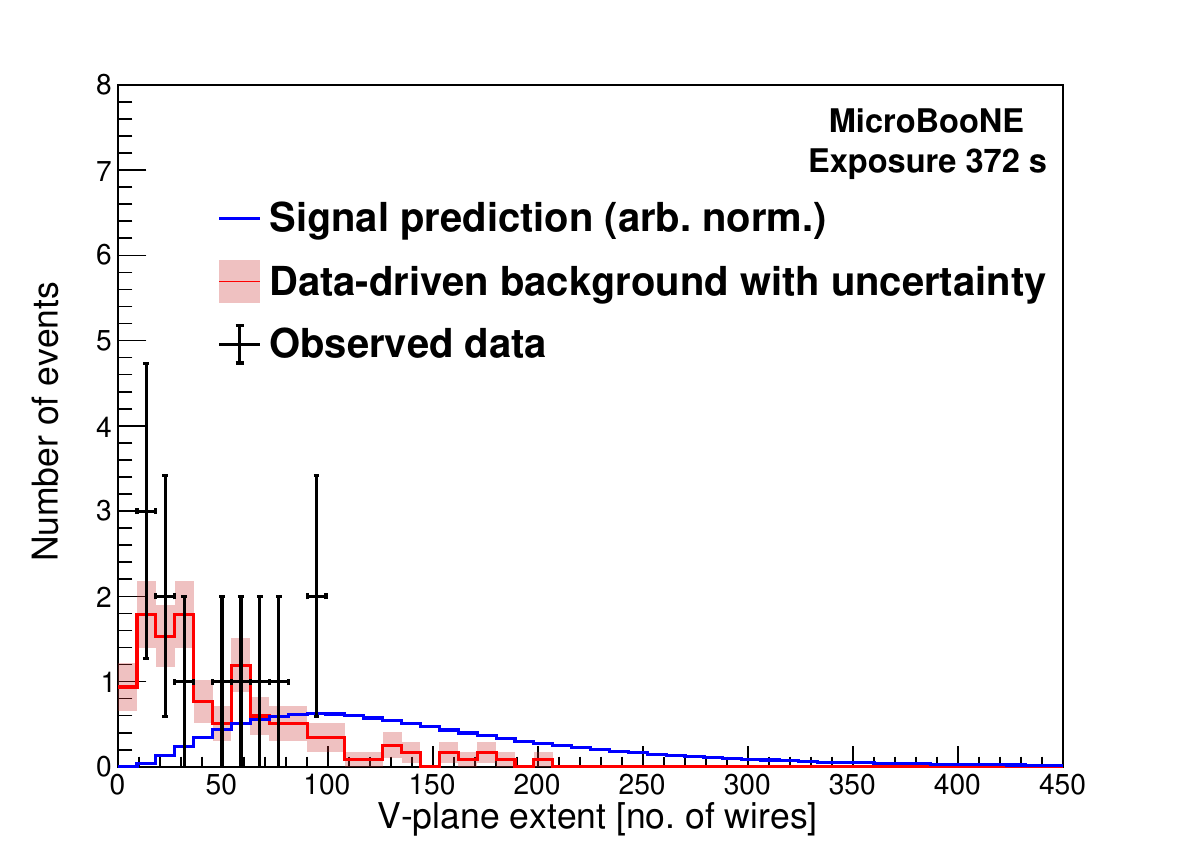}
\\
\includegraphics[width=.49\textwidth]{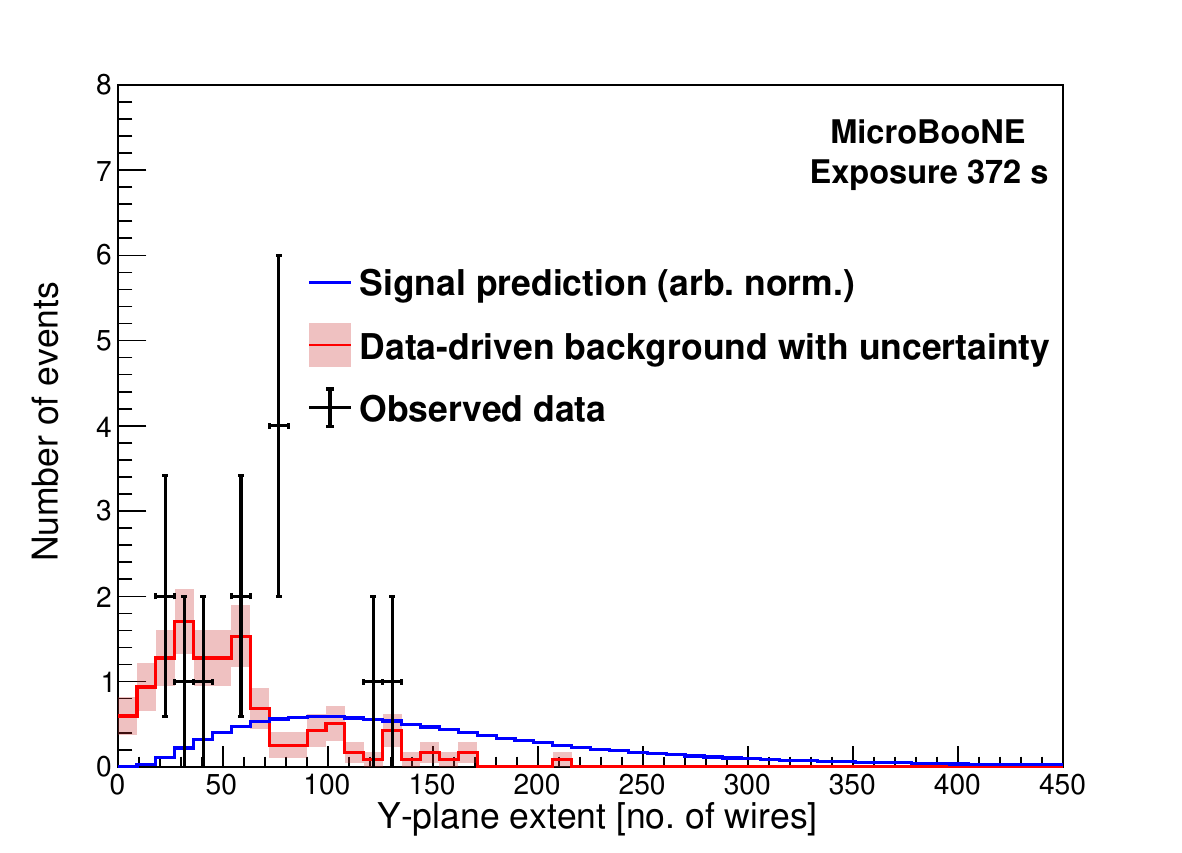}
\includegraphics[width=.49\textwidth]{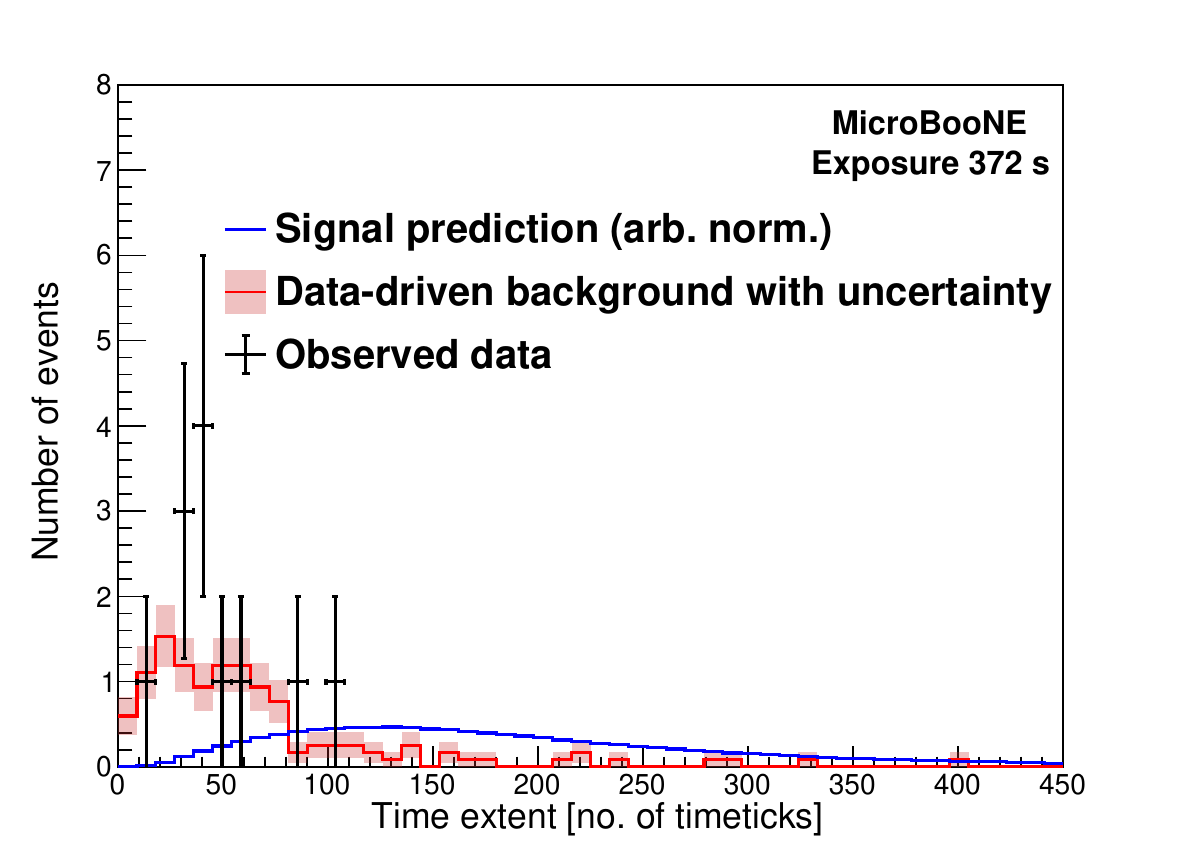}

\caption{
The distributions of $U$,$V$,$Y$-plane and time-tick extents for the signal $n\rightarrow\bar{n}$ (blue) and background (red) events after the CNN score cut are shown. The data points corresponding to $372\,$s of exposure are shown (after unblinding) in black along with statistical uncertainty. The background events, generated with a test sample, are normalized to match the data exposure of $372\,$s, whereas the signal events, for which the clusters are simulated and overlaid onto the background clusters, are arbitrarily normalized as they cannot be precisely scaled to match the data exposure. The samples used to obtain background prediction and data are assumed to have a negligible signal.}
\label{fig:1d_uextent}
\end{figure*}

\begin{table}
\centering
\caption{\label{tab:test_combined_efficiency}The number of predicted signal and background events in the test sample before and after each of the three selection stages.}
\smallskip
\begin{tabular}{l|c|c}
\hline
Selection Stage & Signal & Background\\
\hline
  No selection & 1,633,525 & 1,618,827 \\
   Stage 1 &1,411,164 & 139,802 \\
   Stage 2 & 1,202,281 & 142 \\
   Stage 3 & 1,147,157 & 32 \\
   \hline
   Signal selection efficiency & 70.22\% & 
   - \\
   \hline
   Background efficiency & - & 0.0020\% \\
   \hline
   
\end{tabular}
\end{table}

\section{Systematic uncertainties}
\label{sec:uncs}
The systematic uncertainties on signal and background events are assessed independently. Systematic uncertainties on the signal selection efficiency include contributions from GENIE, Geant4, and detector model variations.

\subsection{GENIE systematics}\label{ub:sys_genie}
The default GENIE nuclear model configuration used in MicroBooNE to simulate \nnbar interactions is the hA-Local Fermi Gas (hA-LFG)~\cite{GENIE:2021npt}. The signal efficiency using simulations with other possible model variations has been evaluated. GENIE offers various models to describe the energy and momentum of the initial state nucleon, such as Bodek-Ritchie (BR)~\cite{Bodek:1980ar} or Local Fermi Gas (LFG). Similarly, final state interactions (FSI) are described in GENIE either through a full cascade model (hN) or an effective model that parameterizes FSI as a single interaction (hA)~\cite{GENIE:2021npt}. For each variation, a new independent signal sample is generated, and the entire selection, as described in section~\ref{sel}, is applied to each of them to evaluate signal selection efficiency, and subsequently, the associated uncertainty. \hbox{Table~\ref{tab:ub_geniesys}} shows the quantitative estimate of uncertainty due to various GENIE models on signal selection efficiency. The fractional uncertainty on the signal selection efficiency, $\eta$, is the uncertainty on the efficiency for each model ($\varepsilon$) with respect to the nominal GENIE hA-LFG model ($\varepsilon_\mathrm{nom}$), defined as
\begin{equation}
\eta = \frac{\varepsilon_\mathrm{nom}-\varepsilon}{\varepsilon_\mathrm{nom}}.
  \label{eq:etagenie}
\end{equation}
This equation does not consider statistical uncertainty on the efficiency evaluated for each model, which is found to be negligible ($2\times 10^{-4}$). The total fractional uncertainty on the signal efficiency due to GENIE systematic uncertainties is estimated to be 4.85\%.




 
\begin{table}
\centering
\caption{\label{tab:ub_geniesys}The fractional uncertainty in signal efficiency $\eta$ is shown for various samples with different GENIE models. The total uncertainty due to GENIE modeling, obtained by taking the squared sum of $\eta$, is estimated to be 4.85\%.}
\smallskip
\begin{tabular}{l|c}  
\hline
    GENIE model &  $\eta$ (\%) \\
     \hline

    hA-BR & 1.17 \\
    hN-BR & 4.56 \\
    hN-LFG & 1.14 \\
      
    \hline
    Total  &  4.85\\
    \hline
    \end{tabular}
\end{table}

\subsection{Geant4 systematics}
Uncertainty from Geant4 accounts for hadron-$^{40}$Ar reinteraction uncertainties. Charged hadrons can interact with external $^{40}$Ar nuclei while traveling through the liquid argon volume.
Inelastic reinteractions of charged hadrons ($\pi^{+}, \pi^{-}, p$) in the LAr volume are simulated by Geant4, and the cross sections of these hadronic reinteractions are varied to account for the corresponding systematic uncertainty. The uncertainty of these scattering processes of protons and charged pions could be significant, especially when there are many charged hadrons in the final state, such as in \nnbar interactions. The impact of hadron reinteraction uncertainty on \nnbar signal efficiency has been evaluated using an event reweighting scheme~\cite{g4_rw}. The systematic uncertainty ($\sigma$) due to charged hadron ($\pi^{+}, \pi^{-}, p$) reinteractions is assessed using the following equation for each hadron,
\begin{equation}
   \sigma = \frac{1}{N_{\textrm{w}}} \sum_{i=1}^{N_{\textrm{w}}}(W_{\textrm{i}}-N)^{2}, 
\end{equation}
 where $i$ runs over the number of  re-weights ($N_{\textrm{w}}$=$1000$) generated for each of the $\pi^{+}$, $\pi^{-}$ and $p$ reinteractions. $W_{\textrm{i}}$ represents the weighted sample which takes weights generated for charged hadrons into account and $N$ represents the nominal sample. Table~\ref{tab:ub_geantsys} shows the fractional uncertainty on the signal efficiency due to hadron reinteraction uncertainties with a total Geant4 uncertainty evaluated to be $2.32\%$.\\

\begin{table}[htbp]
\centering
\caption{\label{tab:ub_geantsys}The fractional uncertainty in signal efficiency $\sigma$ is shown for various samples with different Geant4 reinteraction weights in the last column. The total uncertainty due to Geant4 modeling, obtained by taking the squared sum of each $\sigma$, is estimated to be 2.32\%.}
\smallskip
\begin{tabular}{l|c} 
\hline
    Geant4 reinteractions  & $\sigma$ (\%) \\
     \hline
     $\pi^{+}$ & 0.89 \\
    $\pi^{-}$  & 1.3 \\
    proton & 1.7 \\
      
    \hline
    Total  & 2.32\\ 
    \hline
    \end{tabular}
\end{table}

\subsection{Detector systematics}

 The detector modeling and response uncertainties are evaluated for the signal sample using a novel data-driven technique~\cite{MicroBooNE:2021roa} to account for discrepancies between data and simulation in charge and light response. This uses in situ} measurements of distortions in the TPC wire readout signals due to various detector effects, such as diffusion, electron drift lifetime, electric field, and electronics response, to parametrize these effects at the TPC wire level. 

For each variation, a new independent signal MC sample is generated. The final selection is applied to each of these samples and signal efficiency is calculated. Table~\ref{tab:ub_detsys} shows the fractional uncertainty due to various detector variations on the signal selection efficiency. The fractional uncertainty on signal selection efficiency (quoted in the last column) includes a statistical uncertainty in efficiency, $\eta_{\textrm{err}}$, and uncertainty in efficiency due to each detector variation with respect to the nominal, $\eta_{\textrm{errnom}}$, which are defined as 

\begin{equation}
   \eta_{\textrm{err}} = \sqrt{\frac{\epsilon(1-\epsilon)}{N}},\\
\end{equation}
where $\epsilon$ and $N$ are the signal efficiency and the number of generated events, respectively, for any given model, and 
\begin{equation}
   \eta_{\textrm{errnom}} = \frac{\epsilon_{\textrm{nom}}-\epsilon}{\epsilon_{\textrm{nom}}},
\end{equation}
where $\epsilon_{\textrm{nom}}$ represents the signal efficiency with the nominal sample. The total fractional uncertainty due to detector modeling is evaluated to be 6.72\%.

The total fractional uncertainty on the signal selection efficiency when treating GENIE, Geant4, and detector systematics as being uncorrelated is $8.61\%$. The systematic uncertainty on the background is $17.68\%$, and it corresponds to the statistical uncertainty on the number of final selected background events in the test sample shown in Table~\ref{tab:test_combined_efficiency}. 

\begin{table}[htbp]
\centering
\caption{\label{tab:ub_detsys}The percent uncertainty in signal efficiency, $\eta$, is shown for various samples with different detector systematic variations in the last column. The total uncertainty due to detector systematics, obtained by taking the squared sum of the last column, is estimated to be 6.72\%.}
\smallskip
\begin{tabular}{l|c|c|c}  
\hline
Detector variation & $\eta_{\textrm{err}}$ \%&  $\eta_{\textrm{errNom}}$ \% & $\eta$ \% \\
    \hline
    Recombination &  0.13 & 0.53 & 0.54 \\
    Light yield  & 0.22 & 1.15 & 1.17 \\
     Space charge effect & 0.12 & 0.13 & 0.18\\
     TPC waveform modeling & 0.24 & 6.59 & 6.59 \\
\hline
  Total & & & 6.72\\
\hline
    \end{tabular}
\end{table}

\section{Sensitivity evaluation}
\label{sec:sens}
The final event selection, as described in section~\ref{sel}, yields an expected background of $3.20\pm 1.79\textrm{(stat.)}\pm 0.57\textrm{(syst.)}$ events corresponding to $372\,$s ($3.32 \times10^{26}\,$neutron-years) of exposure, obtained by normalizing the background events reported in table~\ref{tab:test_combined_efficiency} by a factor of 0.1 to predict the background from the data-sized sample. A sensitivity to the intranuclear $n\rightarrow\bar{n}$ lifetime in $^{40}$Ar is evaluated from this exposure using the TRolke statistical method, following a frequentist approach and accounting for both statistical and systematic uncertainties on the background and signal efficiency~\cite{Rolke}. This sensitivity is evaluated assuming the absence of any signal contribution in the sample used for data-driven background determination, treating any observed events as indistinguishable from the background events. The resulting $\tau_m$ sensitivity for $^{40}$Ar corresponds to $6.0\times10^{25}\,$years at $90\%\,$CL.

The DUNE exposure ($1.3 \times10^{35}\,$neutron-years) is projected to be $10^{9}$ orders of magnitude larger than MicroBooNE's. If similar signal selection efficiency can be obtained with DUNE as for MicroBooNE, while maintaining the atmospheric neutrino background rejection reported by DUNE~\cite{Hewes:2017xtr}, then DUNE's statistical-only sensitivity would increase seven-fold.

\section{Fake-data analysis}
\label{sec:fakedata}
The analysis is developed as a blind analysis and the final selection is tested on a dedicated fake-data sample before looking at the data sample reserved for making the final measurement. The fake-data sample corresponds to a data-sized sample of unbiased, off-beam data events ($372\,$s of exposure), which is statistically independent of the data sample and is prepared with a blinded fraction of $x$\% injected $n\rightarrow\bar{n}$ signal, where $x$\% is unknown to the analyzer. As part of the fake-data test, the $x$\% is estimated from the developed analysis framework by performing a fit to the fake data. The final selection, as described in section~\ref{sel}, is applied to the fake-data sample. Out of 158,681 events, 268 events passed the selection, with an expected background of 3.20. 

\begin{figure}[htbp]
\centering
\includegraphics[width=0.99\textwidth]{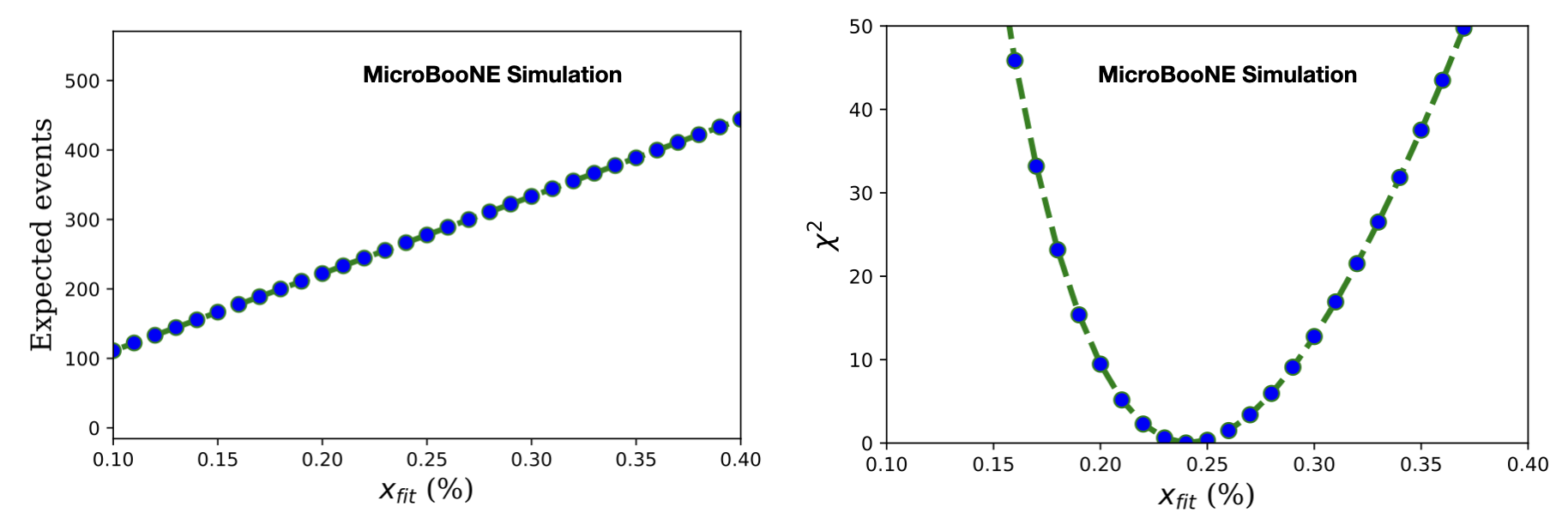}

\caption{Distributions of expected events (left) and $\chi^{2}$ (right) of fake-data observation are shown as a function of the fraction of injected \nnbar events in the fake-data sample, $x_{\textrm{fit}}$.}
\label{fig:fakdedata_chi}
\end{figure}

Next, the compatibility of the fake-data observation with the expectation was quantized by constructing a $\chi^{2}$ as follows:
\begin{equation}
    \chi^{2} = \frac{(O-E)^{2}}{E},
\end{equation}
where $O=268$ is the observed number of events in the fake-data sample, and $E$ is the expected background plus \nnbar signal events and is defined as
\begin{equation}
E = x_{\textrm{fit}}N_{\textrm{g}} \epsilon_{\textrm{s}} + (1-x_{\textrm{fit}})N_{\textrm{g}} \epsilon_{\textrm{b}},
\end{equation}
where $x_{\textrm{fit}}$ is the assumed fraction of injected \nnbar events in the fake-data sample, $N_{\textrm{g}}=158,681$ is the number of events in the fake-data sample, $\epsilon_{\textrm{s}}=0.70$ is the signal selection efficiency, and $\epsilon_{\textrm{b}}=1.97 \times 10^{-5}$ is the background efficiency. $x_{\textrm{fit}}$ is varied to obtain the minimum $\chi^{\textrm{2}}$ value, corresponding to the best-fit $x_{\textrm{bf}}$.
Figure~\ref{fig:fakdedata_chi} shows the expected number of events and $\chi^{2}$ distribution as a function of $x_{\textrm{fit}}$. The best-fit fraction of \nnbar signal is found to be $0.23\%$, whereas the actual fraction revealed after this measurement was performed is 0.25\%. The estimated fraction matches the actual fraction within the 
$1\sigma$ uncertainty of $0.03\%$ demonstrating the overall validity of the selection methods and associated image-based analysis. 

\section{Results}
\begin{figure}[htbp]
\centering 
\includegraphics[width=0.99\textwidth]{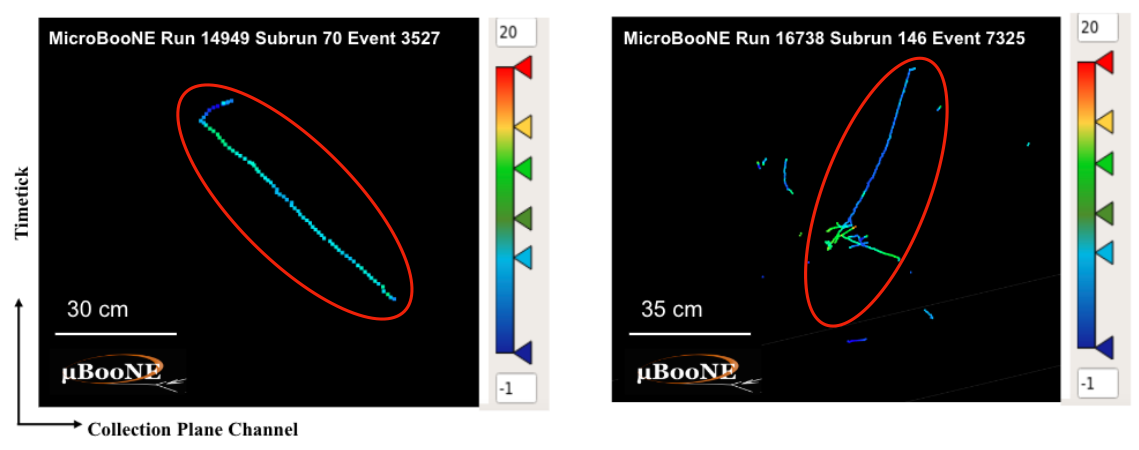}
\caption{Event displays of the two data events (showing the selected interaction cluster) that pass the final analysis selection. Only the selected cluster from the final selection is shown for both events. The $x$-axis represents the collection plane channels and the $y$-axis represents the time-ticks. Color represents the amount of deposited charge in units of ADC values.
}
\label{fig:data_evds}
\end{figure}

In this section, a ``demonstrative'' experimental lower bound on the intranuclear $n\rightarrow\bar{n}$ lifetime in $^{40}$Ar is evaluated at $90\%\,$CL. Unlike the sensitivity which is evaluated in section~\ref{sec:sens}, the ``demonstrative'' lower bound calculation does not assume a zero-signal hypothesis and treats the number of observed events differently from the background events. After successfully validating the developed analysis selection using the fake-data sample as shown in section~\ref{sec:fakedata}, the analysis examined the data sample reserved for reporting the final measurement. Upon applying the analysis selection criteria, $2$ events are observed, consistent with an expected background of $3.20\pm 1.79\textrm{(stat.)}\pm 0.57\textrm{(syst.)}$ events. The observed events are shown in figure~\ref{fig:data_evds}. The selected clusters in both of these events are localized to tens of wires and therefore were mistaken as signal events by the network.

The absence of an excess of events above the expected background prediction leads to a ``demonstrative'' lower bound on the intranuclear $n\rightarrow\bar{n}$ lifetime in $^{40}$Ar of $\tau_m \gtrsim 1.1\times10^{26}$ years at $90\%~$CL (evaluated using the TRolke package in ROOT~\cite{Rolke}). Using eq. (\ref{eq:lifecor}) and $R=5.6\times10^{22}$~s$^{-1}$ for $^{40}$Ar~\cite{Barrow:2019viz}, a limit on the free-neutron equivalent $n\rightarrow\bar{n}$ lifetime is derived as $\tau_{\textrm n-\bar{n}} \gtrsim 2.6\times10^{5}\,$s at $90\%~$CL. Note that such a conversion is subject to an additional uncertainty associated with the use of $R$, which is estimated at $\sim20\%$~\cite{Barrow:2019viz}.

\section{Conclusions}

We have developed and validated a novel approach to search for neutron-antineutron transitions in $^{40}$Ar using the MicroBooNE detector and derived a ``demonstrative'' experimental lower limit on \nnbar lifetime with 90\% CL. This methodology, based on state-of-the-art reconstruction tools and deep learning methods specifically tailored to LArTPC experiments showcases the high sensitivity capabilities of LArTPCs for this topologically unique search. As a proof-of-principle demonstration, we make use of the off-beam data from the MicroBooNE detector under the assumption that this data contains negligible signal events, consistent with Super-Kamiokande results~\cite{PhysRevD.103.012008}, and provide a ``demonstrative'' experimental lower limit on the mean intranuclear neutron-antineutron transition time. As expected, the ``demonstrative'' experimental lower limit is far lower compared to those of previous measurements due to limited exposure and a non-competitive detector mass. The purely topologically-based selection achieves a uniquely high signal selection efficiency of $70.0\%$ and a background rejection efficiency of $99.99\%$; the former of these represents a large improvement over previous results, some of which reported $<10\%$ signal efficiency~\cite{PhysRevD.103.012008}. With an already well-developed methodology, this study demonstrates the future potential of enhanced sensitivities within forthcoming LArTPC-based detectors such as DUNE as mentioned in section~\ref{sec:sens}, in their searches for such rare signals. Further improvements, such as delineating the actual kinematics of signals and backgrounds, along with integration of particle identification, show still more promise. It is important to note that the backgrounds in DUNE and MicroBooNE are distinct. While cosmic ray muons are the dominant backgrounds in MicroBooNE on the surface, atmospheric neutrino interactions are expected to be the main source of backgrounds in DUNE. Nonetheless, the presented analysis demonstrates the usefulness of machine learning techniques, particularly when applied to simple topological extent variables. These variables are unique to LArTPCs due to their fine spatial resolution. The demonstration presented in this paper confirms the capabilities of larger, well-shielded LArTPCs such as DUNE in performing future high-sensitivity searches for baryon number violation.


\acknowledgments
This document was prepared by the MicroBooNE collaboration using the
resources of the Fermi National Accelerator Laboratory (Fermilab), a
U.S. Department of Energy, Office of Science, HEP User Facility.
Fermilab is managed by Fermi Research Alliance, LLC (FRA), acting
under Contract No. DE-AC02-07CH11359.  MicroBooNE is supported by the
following: the U.S. Department of Energy, Office of Science, Offices of High Energy Physics and Nuclear Physics; the U.S. National Science Foundation; 
the Swiss National Science Foundation; 
the Science and Technology Facilities Council (STFC), part of the United Kingdom Research and Innovation; 
the Royal Society (United Kingdom); 
the UK Research and Innovation (UKRI) Future Leaders Fellowship; 
and the NSF AI Institute for Artificial Intelligence and Fundamental Interactions. 
Additional support for 
the laser calibration system and cosmic ray tagger was provided by the 
Albert Einstein Center for Fundamental Physics, Bern, Switzerland. We 
also acknowledge the contributions of technical and scientific staff 
to the design, construction, and operation of the MicroBooNE detector 
as well as the contributions of past collaborators to the development 
of MicroBooNE analyses, without whom this work would not have been 
possible. We gratefully acknowledge W.~J.~Willis for his leadership on the MicroBooNE experiment and formative contributions to $n\rightarrow\bar{n}$ searches with LArTPCs. For the purpose of open access, the authors have applied 
a Creative Commons Attribution (CC BY) public copyright license to 
any Author Accepted Manuscript version arising from this submission.



\bibliographystyle{JHEP}
\bibliography{biblio.bib}

\end{document}